\documentclass[12pt,a4paper,twoside]{article}
\usepackage{epsfig}
\usepackage{baltlat1}
\usepackage{wrapfig}
\begin{document}
\ \
\vspace{0.5mm}

\setcounter{page}{1}
\vspace{8mm}

\titlehead{Baltic Astronomy, vol.~12, XXX--XXX, 2003.}

\renewcommand{\thefootnote}{\fnsymbol{footnote}}%

\titleb{LONG-TERM EVOLUTION OF FU\,ORI-TYPE STARS\\ AT
  INFRARED WAVELENGTHS\footnote[1]{Based on observations with ISO, an ESA project
    with instruments funded by ESA member states (especially the PI countries
    France, Germany, the Netherlands and the United Kingdom) with
    participation of ISAS and NASA.}}

\begin{authorl}
\authorb{\'A.~K\'osp\'al}{1},
\authorb{P.~\'Abrah\'am}{2},
\authorb{Sz.~Csizmadia}{2}
\end{authorl}

\begin{addressl}
  \addressb{1}{Dept.~of Astronomy, E\"otv\"os Lor\'and University, Budapest,
    P.O.~Box 32, H-1518, Hungary}
  
  \addressb{2}{Konkoly Observatory, Budapest, P.O.~Box 67, H-1525, Hungary}
\end{addressl}

\begin{abstract}
  We investigated the brightness evolution of 7 FU\,Ori systems in the
  $1-200\,\mu$m wavelength range using observations from the {\it Infrared
  Space Observatory} (ISO), 2MASS and MSX data. The SEDs were compared with
  earlier ones derived from the IRAS photometry and ground-based observations
  around the epoch 1983.
\end{abstract}

\begin{keywords}
  stars: pre-main sequence -- stars: circumstellar matter -- infrared: stars
  -- stars: individual: V1057\,Cyg
\end{keywords}

\resthead{Evolution of FU\,Ori-type Stars at Infrared
  Wavelengths}{\'A.~K\'osp\'al et al.}



\sectionb{1}{INTRODUCTION}

FU\,Orionis objects are low mass pre-main sequence stars undergoing outburst
in optical light of 5 mag or more. The fading phase after the outburst is well
documented in the optical/near-infrared, but no data have been available so
far at far-infrared wavelengths. The ISO ($1995-98$), provided new photometric
data in the $4.8-200\,\mu$m range: 5 FU Ori-type stars were observed with
ISOPHOT, the infrared photometer on-board ISO. The goal of our study is to
compare far-infrared SEDs based on IRAS photometry ($1983$) with SEDs compiled
from ISOPHOT, MSX, and 2MASS ($1996-2000$), and check for long-time variations of
the infrared flux during a period of 15
years.

\sectionb{2}{OBSERVATIONS AND DATA REDUCTION}

Table 1 lists the infrared photometric data used in this study. We compiled a
list of all confirmed/candidate FU Orionis objects from the literature and
selected those 7 objects for further study where sufficient data were available
at both epochs ($1983$ and $1996-2000$) to create a complete mid/far-infrared
SED.

\begin{center}
\vbox{\footnotesize
\begin{tabular}{llll}
\multicolumn{4}{c}{\parbox{120mm}{\baselineskip=9pt
{\smallbf\ \ Table 1.}{\small\ Overview of infrared photometric data of FU Ori objects.}}}\\
\tablerule
Instrument & Wavelengths [$\mu$m] & Aperture   &  Date \\ 
\tablerule
ground-based & J H K L M N Q                          & $\leq 6''$      & 1970s -- \\ 
IRAS         & 12, 25, 60, 100                        & $1-3'$          & 1983       \\ 
MSX          & 4.25, 4.29, 8.28, 12.13, 14.65, 21.34  & $18''$          & 1996 -- 1997 \\ 
2MASS        & J, H, Ks                               &                 & 1997 -- 2001 \\ 
ISOPHOT      & 4.8, 12, 25, 60, 100, 120, 200         & $43''-180''$    & 1995 -- 1998 \\
\tablerule
\end{tabular}}
\end{center}

The ISOPHOT data reduction was performed using the ISOPHOT Interactive Analysis
Software Package V10.0 (PIA, Gabriel et al. 1997) following a standard scheme
(for details see Abraham et al.) As an error estimate we adopted an absolute calibration uncertainty of 25\%,
which represents well the sum of the random and systematic uncertainties. Colour
corrections were applied for each measurement by convolving the observed SED
with the ISOPHOT filter profiles in an iterative way.

\sectionb{3}{RESULTS}

\begin{figure}
\vskip2mm
\centerline{\psfig{figure=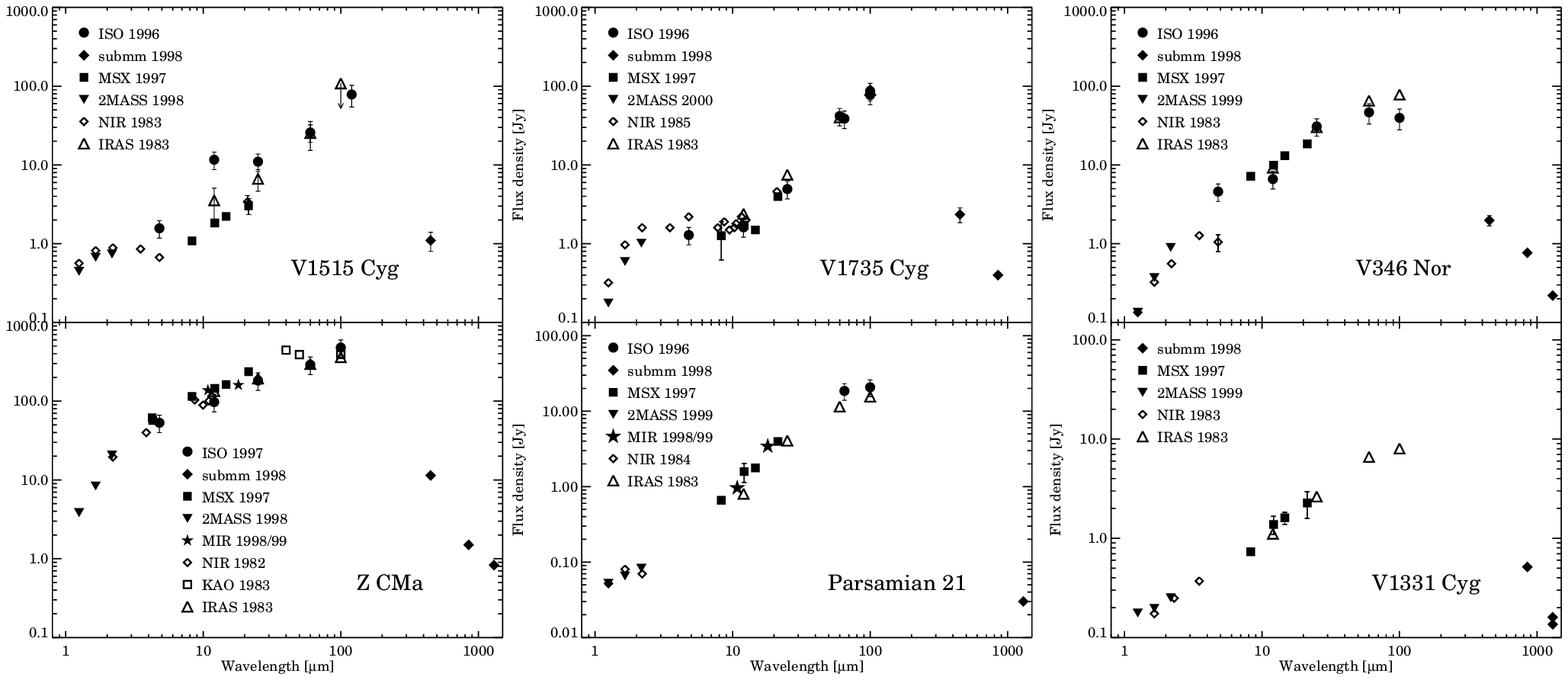,width=126truemm,angle=0,clip=}}
\captionb{1}{SEDs of FU Orionis objects at the two different epochs ($1983$
  with open symbols and $1996-2000$ with filled symbols).}
\vskip2mm
\end{figure}

Our main results, the comparison of SEDs at the two different epochs, are
summarized in Figure 1 and Figure 2. From the figures one can draw the following
conclusions:

\noindent {\bf Near-IR (\boldmath $\lambda \leq 5\, \mu$m):} the sources show various
trends: Parsamian 21, V1331 Cyg and Z CMa are unchanged, V1057 Cyg, V1515 Cyg and
V1735 Cyg have faded, V346 Nor have become slightly brighter.

\noindent {\bf Mid-IR (\boldmath $5\leq \lambda \leq 20\, \mu$m):} only V1057 Cyg shows systematic flux
change: it faded by a factor of 2 during the period.

\noindent {\bf Far-IR (\boldmath $\lambda \geq 60\, \mu$m):}  the majority of the sources does not show
any variation within the measurement uncertainties. The only possible exception
is V346\,Nor exhibiting $2\,\sigma$ drop at $100\,\mu$m. (For V1331\,Cyg there
are no FIR data other than IRAS.)

\sectionb{4}{DISCUSSION: THE CASE OF V1057\,CYG}
We interpret the V1057\,Cyg observations in the framework of the
accretion disk model of Kenyon \& Hartmann (1991, hereafter KH).

\begin{wrapfigure}{0pt}{49mm}
\centerline{\psfig{figure=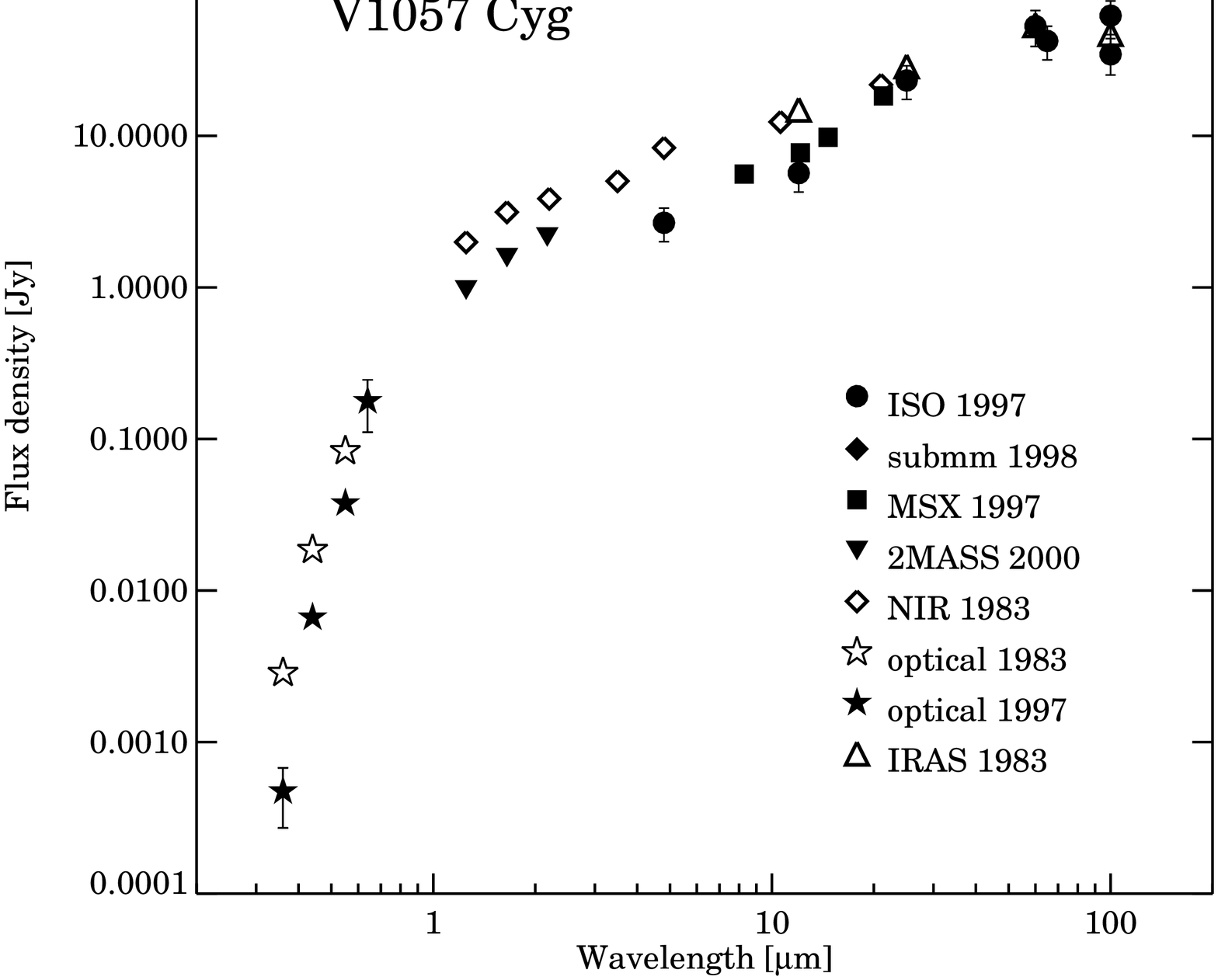,width=48truemm,angle=90,clip=}}
\captionb{2}{SED of V1057\,Cyg at the two different epochs ($1983$
  with open symbols and $1996-2000$ with filled symbols).}
\end{wrapfigure}

\noindent {\bf At \boldmath $\lambda \leq 5\, \mu$m:}
according to KH between $1971$ and $1983$ a general decay of the flux density was
observed, with the lowest amplitude at $5\,\mu$m. The emission comes from
the inner hot part of the accretion disk, and the flux decrease is due to
the dropping accretion rate after the outburst.

The flux decay was continuing between $1983$ and $1997$.
However, no clear wavelength dependence can be observed in this period,
suggesting that the energy budget of the inner part of the disk is now
dominated by reprocessed light rather than by the release of accretion
energy.

\noindent {\bf At \boldmath $5 \leq \lambda \leq 20\, \mu$m:}
according to KH between $1971$ and $1983$ a flux decay was observed which was
wavelength independent and followed the same rate as in the optical. The
emission probably originates from an extended envelope which reprocesses the
optical radiation of the central source.

In the period $1983-1997$ the wavelength independent fading
was still going on and its rate remained synchronised with the rate of the
optical decay (about a factor of 2 at all wavelengths). The results continue
supporting the envelope model.

\noindent {\bf At \boldmath $\lambda \geq 20\, \mu$m:}
according to KH in the lack of real measurements it was assumed that the
far-infrared emission originates also from the envelope and its radiation
drops with time due to the decreasing irradiation from the central source.

Our data however show no flux variation at all. This result suggests
that either the far-infrared flux comes from a different cold component of
the system whose energy budget is independent of the central source; or the
temperature of the outer part of the envelope -- where the long wavelength
emission may come from -- follows the change of irradiance only on a
significantly longer timescale than the 15 years period we covered.

\vskip7mm

ACKNOWLEDGMENTS.\ The ISOPHOT data presented in this paper were reduced using
the ISOPHOT Interactive Analisys package PIA, which is a joint development by
the ESA Astrophysics Division and the ISOPHOT Consortium, lead by the
Max-Planck-Institut f\"ur Astronomie (MPIA).  The work was partly supported by
the grant OTKA T\,037508 of the Hungarian Scientific Research Fund. P.\'A.
acknowledges the support of the Bolyai Fellowship.
\goodbreak

\References

Gabriel, C.\, et al.\, 1999, in {\smalit The ISOPHOT Interactive Analysis PIA, a
  calibration and scientific analysis tool}, Proceedings of the ADASS VI
  conference, ASP Conf.\, Ser.\, 125, p.\,108

Kenyon, S.J., Hartmann, L.\, 1991, in {\smalit The dusty envelopes of FU
  Orionis variables}, ApJ, 383, 664

\end{document}